\documentclass[aps,prl,twocolumn,superscriptaddress,showpacs]{revtex4-1}

\usepackage{longtable}
\usepackage{graphicx}
\usepackage{subfigure}
\usepackage{dcolumn}
\usepackage{bm}
\usepackage{amsmath}
\usepackage[psamsfonts]{amssymb}
\usepackage{lineno}
\usepackage{color}

\begin{document}

\title{Mott transition in granular aluminum}

\author{N. Bachar}
\email[]{nimrodb7@post.tau.ac.il}
\affiliation{Raymond and Beverly Sackler School of Physics and Astronomy, Tel Aviv University, Tel Aviv, 69978, Israel}

\author{S. Lerer}
\affiliation{Raymond and Beverly Sackler School of Physics and Astronomy, Tel Aviv University, Tel Aviv, 69978, Israel}

\author{S. Hacohen-Gourgy}
\affiliation{Raymond and Beverly Sackler School of Physics and Astronomy, Tel Aviv University, Tel Aviv, 69978, Israel}

\author{B. Almog}
\affiliation{Raymond and Beverly Sackler School of Physics and Astronomy, Tel Aviv University, Tel Aviv, 69978, Israel}

\author{H. Saadaoui}
\affiliation{Laboratory for Muon Spin Spectroscopy, Paul Scherrer Institute, 5232 Villigen PSI, Switzerland}

\author{E. Morenzoni}
\affiliation{Laboratory for Muon Spin Spectroscopy, Paul Scherrer Institute, 5232 Villigen PSI, Switzerland}

\author{G. Deutscher}
\affiliation{Raymond and Beverly Sackler School of Physics and Astronomy, Tel Aviv University, Tel Aviv, 69978, Israel}

\date{\today}

\begin{abstract}
The presence of free spins in granular Al films is directly demonstrated by $\mu$SR measurements. A Mott transition is observed by probing the increase of the spin-flip scattering rate of conduction electrons as the nano-size metallic grains are being progressively decoupled. Analysis of the magneto-resistance in terms of an effective Fermi energy shows that it becomes of the order of the grains electrostatic charging energy at a room temperature resistivity $\rho \approx 50,000~\mu\Omega~cm$, at which a metal to insulator transition is known to exist. As this transition is approached the magneto-resistance exhibits a Heavy-Fermion like behavior, consistent with an increased electron effective mass.
\end{abstract}



\maketitle

Thanks to advances in the development of the Density functional Mean Field Theory (DMFT)~\cite{Georges1996}, considerable advances have been made in recent years towards a detailed understanding of the Mott metal to insulator transition, predicted to occur when the electron-electron interaction is of the order of the bandwidth~\cite{Mott1949,*Mott1968}. However the experimental observation of this transition has remained a challenge in three dimensional systems. This is because in a homogeneous metal the Coulomb interaction is by several orders of magnitude smaller than the bandwidth, even in the presence of a relatively high concentration of impurities~\cite{Mott1974,Imada1998}.

\par

We show here that a Mott transition takes place in granular metals, as nano-size grains are being decoupled from each other by a progressive reduction of the inter-grain tunneling probability. Two of the main features of this transition predicted by DMFT theory, an increase of the electron effective mass and a non-critical behavior of the electronic density of states as the transition is approached, have been observed.

\par

These observations have been made possible by the presence of free spins in granular Aluminum films, which we confirm here by direct $\mu$SR measurements. Interaction of these spins with conduction electrons results in a negative magneto-resistance~\cite{Beal-Monod1968}. We have used it as a tool to follow changes of the effective Fermi energy of the granular medium as the transition is being approached. When it occurs, at a room temperature resistivity of about 50,000~$\mu\Omega cm$, we find that the effective Fermi energy is of the order of the grain's charging electrostatic energy. The superconducting critical temperature of the films remains relatively high up to close to the transition, indicating there is no drastic reduction of the density of states up to the transition.

\par

\begin{figure}
    \begin{center}
        \includegraphics[width=0.4\textwidth]{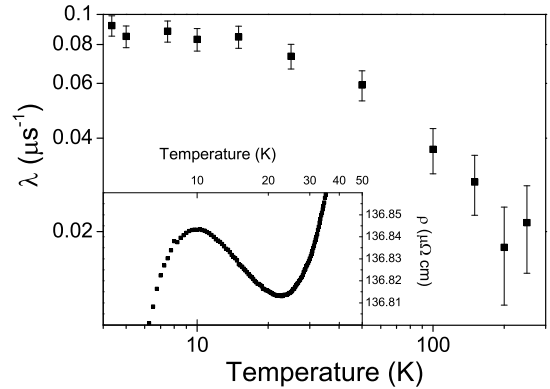}
        \caption{Temperature dependence of the muon spin relaxation rate of electronic origin $\lambda$ for a sample with $\rho_{300K}~\approx~140~\mu\Omega cm$. $\lambda$ appears to saturate around a temperature where $\rho(T)$ starts to increase~\cite{Bachar2013}}
        \label{Fig1}
    \end{center}
\end{figure}

Low energy muon spin rotation/relaxation (LE\_$\mu$SR) experiments~\cite{Morenzoni2004} were performed on film not too close to the MIT transition. The measurements were performed at the Swiss Muon Source on the $\text{{$\mu$}E4}$ beam-line, at the Paul Scherrer Institute, in Switzerland. With implantation energy of 10~keV all the muons stop in the 100~nm film, with a mean range of 68~nm and a stopping width (rms) of 15~nm. The time evolution of the polarization of the muon ensemble implanted in the sample, $P(t)$, measured via detection of the emitted decay positron intensity as a function of time after thermalization, is very sensitive to the local magnetic environment and to the presence of spins. Polarization spectra taken at different temperatures under zero field conditions decay with a small rate, which has two contributions. The first, which is temperature independent, is due to the nuclear moments of Al. In addition a temperature dependent exponential relaxation of electronic origin is observable with rate $\lambda$ (Figure~\ref{Fig1}). This rate reaches a value $\lambda =0.085~\mu{{s}^{-1}}$ at a temperature corresponding to the low temperature increase of the resistivity. The analysis of muon spin rotation measurements on a relatively metallic sample (Figure~\ref{Fig1}) shows that free spins are present at a concentration level of 350 ppm.

The presence of localized magnetic moments in bulk Al is not easily understood. However the granular structure can lead to some interesting and important phenomena. It is well known that magnetic properties are expected in small metallic particles due to the electronic confinement~\cite{Kubo1962}. A small Al grain having an energy level splitting $\delta$ larger than the thermal energy ${{k}_{B}}T$ can be considered as a quantum dot showing a Kondo effect due to spin flip scattering process with the unpaired electron at the highest occupied energy level. This is known as the odd-even effect of Quantum dots~\cite{Cronenwett1998}. Alternately, scattering can occur with spins located at the metal-oxide interface between neighboring grains. Such spins are observed in the flux noise measurements of Al/Al$_{2}$O$_{3}$/Al based SQUID and q-bit devices~\cite{Sendelbach2008,Anton2013}. Considering that these grains have a size of about 2~nm, we calculate a spin interface density of about ${{10}^{16}}~{{m}^{-2}}$. Flux noise and other magneto-metric measurements on Al oxide interface resulted in a surface spin density of approximately $5\times{{10}^{17}}~{{m}^{-2}}$~\cite{Sendelbach2008,Anton2013,Bluhm2009}. A surface spin density of approximately $10^{16}~m^{-2}$ was shown to fit the $1/f$ flux noise observed in superconducting devices Faoro and Ioffe~\cite{Faoro2008}. It is conceivable that the free spin density at interfaces is different of from that at free surfaces~\cite{Lee2014}.

\par

In the resistivity regime covered here the grain size distribution is roughly constant~\cite{Deutscher1972,*Deutscher1973,*Deutscher1973a}. Therefore the spin density is not expected to change markedly whether it reflects the presence of Kubo moments~\cite{Kubo1962}, or the presence of spins at the metal/metal oxide interfaces~\cite{Sendelbach2008,Anton2013}. We therefore retain the concentration of 350 ppm for our further calculations as a constant value throughout the entire phase diagram.

\par

The fabrication of granular Al films has been described elsewhere in detail~\cite{Deutscher1972,*Deutscher1973,*Deutscher1973a}. Briefly, Al is evaporated in the presence of a reduced pressure of oxygen. The films were deposited unto a substrate, held either at room temperature or at liquid nitrogen temperature. They consist of nano-scale grains of crystalized Aluminum separated by thin oxide barriers. All samples studied had a thickness of 100~nm. The respective amounts of Al metallic grains and Al oxide were controlled by the Al evaporation rate and oxygen partial pressure during evaporation. As the oxide volume fraction is increased the grain size first goes down and eventually reaches a constant value of about 2~nm for films deposited at liquid nitrogen temperature, and 3~nm for films deposited at room temperature. This occurs when the room temperature value of the macroscopic resistivity reaches about 100~$\mu\Omega cm$. As the macroscopic resistivity increases,${{T}_{c}}$ rises from that of bulk Al (1.2~K), reaches a rather flat maximum value of 3.2~K (films deposited at liquid nitrogen temperature) or 2.2~K (films deposited at room temperature) for resistivity values ranging from 100 to 300~$\mu\Omega cm$, before slowly going down for resistivities higher than 500~$\mu\Omega cm$. A film deposited at liquid nitrogen temperature having a resistivity of about 10,000~$\mu\Omega cm$ is still superconducting at 1.8~K.

\par

The negative magneto-resistance of a metal containing magnetic impurities is due, at low fields, to the difference in the scattering rates of spin up and spin down electrons. In the presence of a magnetic field oriented along the positive $z$ direction, which tends to orient the magnetic moment of the impurities along this direction, there are more scattering events of a spin down electron into a spin up than that of a spin up electron scattering into a spin down. But the lower density of available states below the Fermi level where spin up electrons end up due to the Zeeman energy gain forbids these scattering processes because of the exclusion principle~\cite{Beal-Monod1968}. The total contribution of the spin-flip scattering amplitudes to the resistivity is decreased. The difference in the scattering rates of the two processes increases as ${{\alpha }^{2}}={{\left( {g{{\mu }_{B}}H}/{{{k}_{B}}T}\; \right)}^{2}}$, and so does the magneto-resistance. In this regime the field dependence of the scattering amplitudes is negligible. The interpretation of the ${{\left( {H}/{T}\; \right)}^{2}}$ dependence of the negative magneto-resistance of these films in terms of the presence of localized magnetic moments, proposed in a previous publication~\cite{Bachar2013}, is confirmed by the present direct observation of free spins.

\par

Here we focus on the large increase of the magneto-resistance seen as Al grains are being progressively decoupled and a metal to insulator transition is approached. In dilute alloys the magneto-resistance is proportional to the magnetic impurity concentration~\cite{Beal-Monod1968}. Since we expect it to remain roughly constant in our films, the magneto-resistance should remain roughly constant as the oxide content and the normal state resistivity increase. However it increases by several orders of magnitude~\cite{Bachar2013}, rising in fact faster than the resistivity itself with a power law $\Delta \rho \propto {{\rho }^{1.38}}$. Since this rapid increase cannot be attributed to an increase of the spin density, the origin of the increasing spin scattering of conduction electrons must rather lie, we believe, in an increase of the scattering amplitudes as the metal to insulator transition is approached.

\par

Since the magneto-resistance results from the difference of the scattering rates of spin up and spin down electrons, we can write it as:
\begin{equation}\label{DeltaRho_Tau}
    \Delta \rho ={{\rho }_{+}}-{{\rho }_{-}}=\frac{{{m}^{*}}}{n{{e}^{2}}}\left( \frac{1}{{{\tau }_{+}}}-\frac{1}{{{\tau }_{-}}} \right)
\end{equation}
The scattering rates are given as:
\begin{equation}\label{Tau_Spin}
    \frac{1}{{{\tau }_{\pm }}}=\frac{k{{m}^{*}}{{v}_{0}}c}{\pi {{\hbar }^{3}}}f\left( V,J,\left\langle {{S}_{z}} \right\rangle ,\left\langle S_{z}^{2} \right\rangle ,S,{{\epsilon }_{\pm }},\alpha  \right)
\end{equation}
where $k$ and ${{m}^{*}}$ are, respectively, the wave number and effective mass of the conduction electron, ${{v}_{0}}$ is the atomic volume of the host metal and $c$ is the magnetic impurity concentration in ppm. $f$ is a general function defined in ref.~\cite{Beal-Monod1968} by the Coulomb interaction $V$, the interaction constant $J$, the equilibrium average of the operators ${{S}_{z}}$ and $S_{z}^{2}$ for the spin component in the $z$ direction, the impurity spin $S$, the energy shifts ${{\epsilon }_{\pm }}={{E}_{F}}\mp {{\mu }_{B}}H$ of the spin up/down states and $\alpha =\left( {g{{\mu }_{B}}H}/{{{k}_{B}}T}\; \right)$ with the $g$ value of the impurity.

\par

We assume that the large increase of the magneto-resistance is primarily due to the increase of the scattering amplitudes, due to variations of the carrier density $n$ and of the effective mass ${{m}^{*}}$ as grains are being progressively decoupled:
\begin{equation}\label{DeltaRho_EffMass}
    \Delta \rho \propto \frac{{{m}^{{{*}^{2}}}}}{n}
\end{equation}
Both a decrease in the density of conduction carriers and an increase in their effective mass can contribute to an increase of the magneto-resistance. To evaluate the increase of the effective mass with the normal state resistivity, we use the Hall mobility result~\cite{Bandyopadhyay1982}:
\begin{equation}\label{HallMob}
    {{\mu }_{H}}\propto {{\rho }^{-0.5}}
\end{equation}
Bandyopadhyay et al. have presented their results as a function of the normal state resistivity measured at 4.2~K, but for samples for which $\rho <2\times {{10}^{4}}~\mu \Omega cm$ the temperature dependence of the resistivity is negligible and the Hall carrier concentration varies as a function of $\rho$ in the same way as the mobility does. Since the magneto-resistance varies as ${{\rho }^{1.38}}$, we conclude that:
\begin{equation}\label{EffMass}
    {{m}^{*}}\propto {{\rho }^{0.44}}
\end{equation}
For the cleaner sample measured by Bandyopadhyay et al., with $\rho \approx 230~\mu \Omega cm$, the value of the Hall constant is ${{R}_{H}}={{\mu }_{H}}\cdot \rho =0.47\times {{10}^{-10}}~{\Omega m}/{T}\;$, getting close to the bulk Al value of $0.3\times {{10}^{-10}}{\Omega m}/{T}\;$. Therefore we will admit that in samples having normal state resistivities lower than $\rho \approx 100~\mu \Omega cm$ the basic electronic properties - effective mass and Fermi energy - have their bulk values. According to Eq.~\ref{EffMass}, the electron effective mass increases by a factor 10 in as sample having resistivity of $\rho \approx 10,000~\mu \Omega cm$.

\par

\begin{figure}
    \begin{center}
        \includegraphics[width=0.45\textwidth]{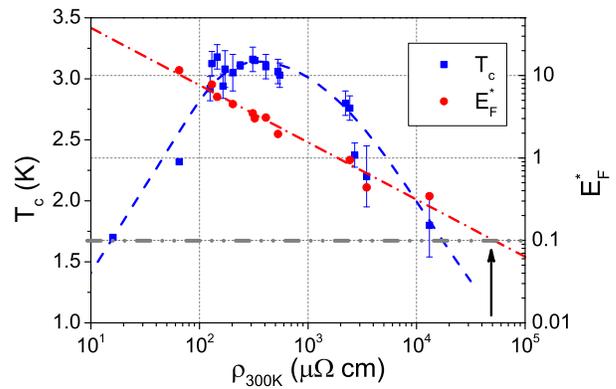}
        \caption{Critical temperature, ${{T}_{c}}$, and effective Fermi energy as a function of the room temperature resistivity. The dashed line is a guide to the eye, the dash dot line follows $E_{F}^{*}\propto {{\rho }^{-0.7}}$ and the dash double dot line is showing the limit of the electrostatic energy for a 2~nm grain. The arrow marker points to the critical resistivity $\rho \approx 50,000~\mu \Omega cm$ at which Mott transition occurs.}
        \label{Fig2}
    \end{center}
\end{figure}

In order to discuss the approach to the metal to insulator transition it is convenient to present Eq. 26 of Ref.~\cite{Beal-Monod1968} in a slightly different form:
\begin{equation}\label{DeltaRho_BMD}
    \Delta \rho =-\beta \frac{\hbar }{{{e}^{2}}}a{{c}_{i}}{{\left( \frac{J}{{{E}_{F}}} \right)}^{2}}{{\alpha }^{2}}u
\end{equation}
where $\beta$ is a number that depends on the atomic packing factor of the material, ${\hbar }/{{{e}^{2}}}\;$ is the quantum resistance, $a$ is the interatomic distance, ${{c}_{i}}$ is the magnetic impurity concentration in ppm, $J$ the interaction constant, ${{E}_{F}}$ the Fermi energy, $\alpha =\left( {g \mu H}/{{{k}_{B}}T}\; \right)$ and $u$ is a function of the spin of the impurity $S$, the Coulomb interaction $V$, $J$ and ${{E}_{F}}$. Figure~\ref{Fig2} shows the variation of the effective Fermi energy as a function of the room temperature resistivity of the films, assuming that all other factors in Eq.~\ref{DeltaRho_BMD} remain constant, $T=20~K$, $H=14~T$ and an impurity concentration of 350 ppm. In order to evaluate the unknown factor $\beta {{J}^{2}}u$ we have assumed that the value of the Fermi energy in the lowest resistivity sample showing the magneto-resistance scaling behavior ($\rho \approx 65~\mu \Omega cm$) is that of the bulk, ${{E}_{F}}=11.6~eV$. This is in keeping with the fact that at this resistivity the Hall constant is unchanged from the bulk value, as noted above.

\par

As can be seen from Fig.~\ref{Fig2}, ${{E}_{F}}\propto {{\rho }^{-0.7}}$. Its extrapolated value hits the electrostatic charging energy $U=85~meV$of the grains at a room temperature resistivity of about $50,000~\mu \Omega cm$,  the value of $U$ being calculated using the grain size of 2 nm and the dielectric constant of $\text{A}{{\text{l}}_{\text{2}}}{{\text{O}}_{\text{3}}}$, $\varepsilon =8.5$. This room temperature resistivity value is known to be that where the low temperature dependence of the resistivity goes from logarithmic to exponential, which is considered as a reliable criterion for the metal to insulator transition~\cite{Deutscher1980}. This is our central result: the metal to insulator transition in nano-scale granular Aluminum occurs when the effective Fermi energy becomes of the order of the electrostatic charging energy of the grains. This strongly suggests that the transition is of the Mott type. Since the density of states varies as ${n}/{{{E}_{F}}}\;$ and since $n\propto {{\rho }^{-0.5}}$, the dependence of the density of states on $\rho$ is quite weak, $N\left( 0 \right)\propto {{\rho }^{0.2}}$. This is consistent with the predicted non-critical behavior of the density of states at the Mott transition~\cite{Georges1996}.

Kawabata has calculated the bandwidth of a system consisting of small metallic particles with an electronic level splitting $\delta $ and inter-grain electron tunneling matrix element $t$. Assuming that there are $z$ adjacent particles, he finds in the weak coupling limit a bandwidth $w\cong {\pi z{{t}^{2}}}/{\delta }\;$~\cite{Kawabata1977}. Since the grain size and therefore $\delta $ should be roughly constant, the control parameter is $t$, which varies with the resistivity as ${{\rho }^{-1}}\propto {{t}^{2}}$~\cite{Beloborodov2007}. As the grain's oxide barrier thickness increases, so does the resistivity, $t$ reduces and as a result $w$ decreases. However, these calculations do not take into account electron-electron interactions.

\par

\begin{figure}
    \begin{center}
        \includegraphics[width=0.45\textwidth]{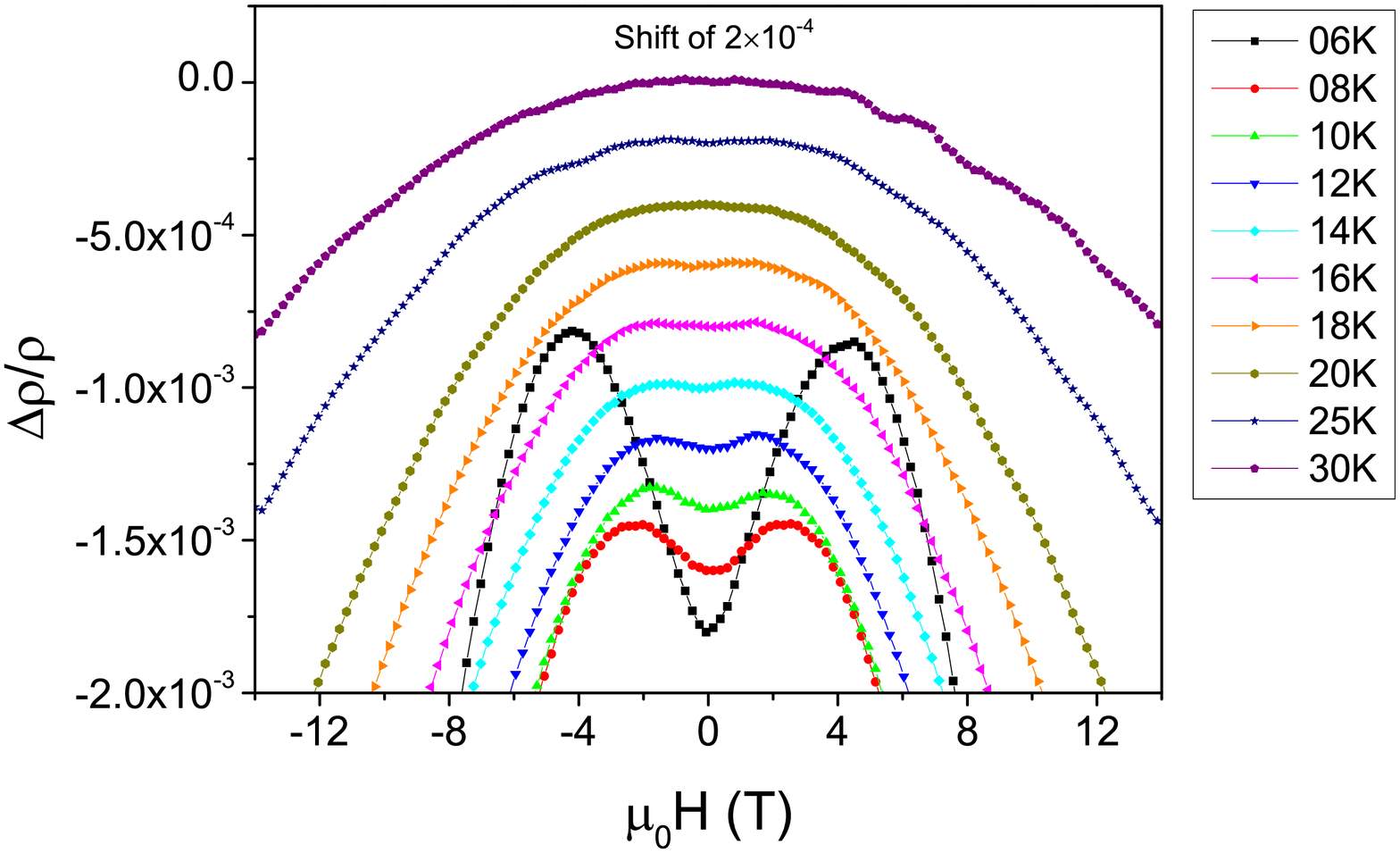}
        \caption{Magneto-resistance of a high resistivity film ($\rho \approx 13,000~\mu \Omega cm$) at low temperatures. Below a temperature ${{T}_{MR}}$ (for this sample ${{T}_{MR}}\approx 20~K$) the magneto-resistance is positive at low fields and negative at high fields. Down to 8~K the variation of the magneto-resistance changes sign at a fixed field, as is the case for a Kondo lattice. Below 8~K that field increases due to superconducting fluctuations (Ghost critical field effect).}
        \label{Fig3}
    \end{center}
\end{figure}

The increasing effective mass in high resistivity films might induce some similarities with the case of Heavy Fermions. Such similarity is indeed observed in their respective magneto-resistance behaviors. In the Kondo diluted case the magneto-resistance remains negative at all temperatures. But in Heavy Fermions there is a change of sign of the magneto-resistance at low temperatures at a fixed field~\cite{Rauchschwalbe1987,Taillefer1988}. In granular Al, the low field magneto-resistance changes sign from negative to positive below a temperature ${{T}_{MR}}$. For high resistivity films ${{T}_{MR}}$ is well above the superconducting critical temperature, hence it is unlikely that the change of sign can be attributed to superconducting fluctuations. For instance, in the most resistive (${{\rho }_{300K}}\approx 13,000~\mu \Omega cm$) superconducting sample measured, which still has a critical temperature of 1.8~K, the low field magneto-resistance turns positive below 20~K, as can be seen in Figure~\ref{Fig3}. Down to $T =8~K$, the field at which the magneto-resistance changes sign remains of the order of 2~T. It is only below 8~K that the magneto-resistance changes sign at a higher field, due to superconducting fluctuations known as the Ghost Critical Field (GCF) effect~\cite{Kapitulnik1985}. Therefore, above 8~K, the change of sign in the magneto-resistance is a property of the normal state where granular Al films show a Heavy-Fermion like behavior.

Finally we would like to address the coexistence of enhanced superconductivity and spin-flip scattering in granular Al. The presence of magnetic moments is known to be detrimental to conventional BCS superconductivity, $T_{c}$ decreases as~\cite{Abrikosov1961,Maki1969}:
\begin{equation}\label{Tc_AGM}
    \ln \left( \frac{{{T}_{c}}}{{{T}_{c0}}} \right)=\psi \left( \frac{1}{2} \right)-\psi \left( \frac{1}{2}+\frac{\hbar }{4\pi {{k}_{B}}{{T}_{c}}{{\tau }_{s}}} \right)
\end{equation}
where ${{T}_{c}}_{0}$ is the critical temperature of an unperturbed superconductor, $\psi \left( x \right)$ is the digamma function, and ${\hbar }/{{{\tau }_{s}}}\;$ is the perturbation strength. The latter is given by~\cite{Ludwig1971}
\begin{equation}\label{Tau_AGM}
    \frac{\hbar }{{{\tau }_{s}}}=\frac{\pi}{4}{c_{i}}S\left( S+1 \right)N\left( 0 \right){{\left| J \right|}^{2}}
\end{equation}
Considering $N\left( 0 \right)=0.381~eV^{-1}$  for aluminum, typical values of $S=0.5$, $J=1~eV$ and a concentration of 350~ppm, we get ${\hbar }/{{{\tau }_{s}}}\;\approx 0.1~meV$. This gives ${{T}_{c}}\approx 2.73~K$ for a superconductor with ${{T}_{c0}}\approx 3.2~K$. It should be noted here that in order to completely quench superconductivity, i.e. ${\hbar }/{{{\tau }_{s}}}\;=\Delta \left( 0 \right)\approx 0.5~meV$~\cite{Cohen1968a}, a magnetic impurity concentration of approximately 1750~ppm is needed, 5 times larger that was found in granular Al film. Nevertheless, further experimental and theoretical work should be done in order to shed more light on the interplay between enhanced superconductivity and magnetism in granular Al films.

\par

In conclusion, the scattering of conduction electrons by free spins, whose presence in granular Al/$\text{A}{{\text{l}}_{2}}{{\text{O}}_{3}}$ films was directly confirmed by muon spin rotation experiments, increases by several orders of magnitude as the grains are progressively decoupled. This large increase is attributed to the decrease in the effective Fermi energy. At high resistivities, the effective Fermi energy is of the order of the electrostatic energy of the grain and a Mott transition is likely to occur. The critical resistivity at this transition is in agreement with previous experimental observations of a metal to insulator transition in granular Al films. A Heavy-Fermion like behavior is observed in the magneto-resistance of the films as the Mott transition is approached and the effective mass increases. Elucidation of the coexistence of free spins in granular Al with enhanced superconductivity requires further work.

\begin{acknowledgments}
Useful discussions with Philippe Nozi\`{e}res are gratefully acknowledged
\end{acknowledgments}

This work was partially supported by EOARD Award FA8655-10-1-3011.

\bibliographystyle{apsrev4-1}
\bibliography{PRL2014_Mott_GranularAl}

\end{document}